\documentclass[english,twocolumn]{revtex4-1}
\usepackage[T1]{fontenc}
\usepackage[latin9]{inputenc}
\usepackage{amstext}
\usepackage{graphicx}
\usepackage{esint}
\usepackage{babel}
\begin{document}
	
	\title{Itinerant chimeras in an adaptive network of pulse-coupled 
	oscillators}
	
	\author{{\normalsize{}Dmitry V. Kasatkin, Vladimir~V. Klinshov and Vladimir
			I. Nekorkin}}
	
	\affiliation{Institute of Applied Physics of the Russian Academy of 
	Sciences,
		46 Ul'yanov Street, 603950, Nizhny Novgorod, Russia}
	
	\begin{abstract}
		In a network of pulse-coupled oscillators with adaptive coupling, we 
		discover a 
		novel dynamical regime which we call an ``itinerant  chimera''. 
		Similarly as in 
		classical chimera states, the network splits into two domains, the 
		coherent and 
		the incoherent ones. The drastic difference is that the composition of 
		the 
		domains is volatile, i.e. the oscillators demonstrate spontaneous 
		switching 
		between the domains. This process can be seen as traveling of the 
		oscillators 
		from one domain to another, or as traveling of the chimera core across 
		the 
		network.  We explore the basic features of the itinerant  chimeras, 
		such as the 
		mean and the variance of the core size, and the oscillators lifetime 
		within the 
		core. We also study the scaling behavior of the system and show that 
		the 
		observed regime is not a finite-size effect but a key feature of the 
		collective 
		dynamics which persists even in large networks.
	\end{abstract}
	
	\maketitle
	
	\section{Introduction}
	
	Networks of interacting nodes are omnipresent in nature and technology
	\cite{Boccaletti2006}. In recent decades, a specific type of collective
	behavior called ``chimera states'' is intensively explored in networks
	of coupled oscillators. Chimera states manifest themselves as spontaneous
	symmetry breaking in systems of identical and symmetrically coupled
	oscillators which split into phase-coherent and incoherent parts.
	First observed by Kuramoto and Battogtokh \cite{Kuramoto2002} and
	later named ``chimeras'' by Abrams and Strogatz \cite{Abrams2004},
	this type of partial synchronization later attracted much attention
	of specialists in dynamical networks. Chimera states were discovered
	and studied for networks of various configurations, and experimental
	observations were provided as well (see the reviews 
	\cite{Panaggio2015,Omelchenko2018}
	and references therein). 
	
	The analytical study of the chimera states was carried out in the continuum 
	limit, see for example \cite{Abrams2008,Wolfrum2011Chaos,Omelchenko2013}. 
	However, for the finite network size the rigorous analysis is hardly 
	possible, 
	and the results rely on the intensive numerical studies. It was shown that 
	finite-size effects have a pronounced influence on the chimera states. In 
	particular, the life-time of chimeras quickly decreases as the number of 
	oscillators in the network becomes smaller \cite{Wolfrum2011PRE}. Another 
	characteristic feature is the Brownian-like motion of the chimera position, 
	i.e. location of the coherent domain in the network \cite{Omelchenko2010}. 
	The 
	effective  diffusion coefficient quickly drops as the network size grows 
	which 
	allows to associate the motion to finite-size effects.
	
	In the present paper, we demonstrate a new type of chimera-like behavior 
	which 
	we call an ``itinerant  chimera''. Similarly with classical chimeras, in 
	this 
	state the network splits into the coherent and the incoherent domains. 
	However, 
	the drastic difference is the volatile composition of the domains. As the 
	time 
	passes, each oscillator demonstrates spontaneous transitions between the 
	domains, so that none of them remains in the same domain forever. From the 
	collective dynamics viewpoint, this process can be seen as the traveling of 
	the 
	synchronized core across the network. Importantly, the core motion is not 
	just 
	a finite-size effect observed for small number of interacting units, but 
	rather 
	a key characteristic of the network dynamics which persists even for large 
	networks. 
	
	\begin{figure}[t]
		\includegraphics[width=85mm]{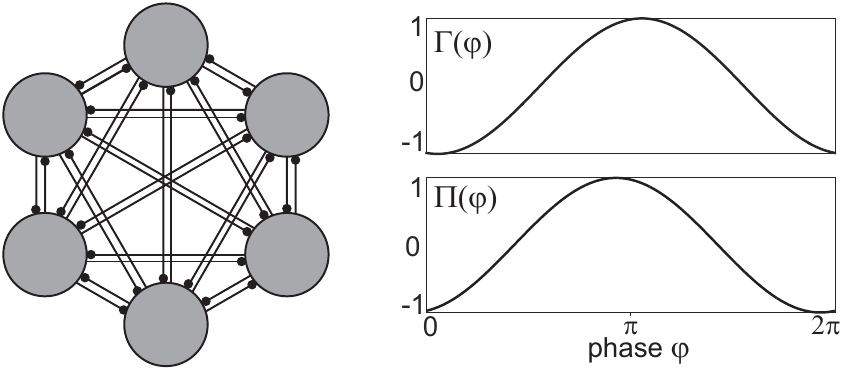}
		
		\caption{\label{fig:sketch}Left: the circuitry of the studied network. 
			Large gray circles denote oscillators, black lines denote 
			all-to-all 
			coupling, small black circles stand for adaptive coupling strength. 
			Right: 
			the plots of the PRC $\Gamma(\varphi)=-\sin(\varphi+\alpha)$ and 
			the 
			adaptivity function $\Pi(\varphi)=\sin(\varphi+\beta)$ for the 
			standard 
			parameter values $\alpha=1.4$ and $\beta=4.94$.}
	\end{figure}
	
	The motion of the chimera's core was reported in a number of previous 
	works. In 
	\cite{Omelchenko2010} it was shown that Brownian-like motion is intrinsic 
	for 
	chimeras, but the effective diffusion coefficient vanishes for large 
	networks. The disrupted chimera ordering with the wandering incoherent 
	domain was observed in a lattice of spins \cite{Dednam2018}.	
	In \cite{Rosin2014} the so-called resurgence of chimera states was reported 
	which manifests itself as spontaneous emergence of chimeras at random 
	positions 
	where they exist for some time and later disappear. Transient chimeras in 
	modular networks were observed in \cite{Shanahan2010} where the synchrony 
	in 
	different modules was rising and falling in irregular manner. Alternating 
	chimeras were observed where the coherent and  incoherent domains swapped 
	on 
	course of the network dynamics  \cite{Haugland2015,Ma2010}.  In 
	\cite{Bick2018} 
	hetoroclinic switching between chimeras was demonstrated which can be 
	interpreted as periodical traveling of the chimera across the network. The 
	most 
	typical type of chimera motion is a constant-speed drift which may be 
	induced 
	by such factors as the sign-alternating coupling function \cite{Xie2014}, 
	coupling asymmetry \cite{Bick2014,Bera2016}, nonlinear coupling 
	\cite{Mishra2017} or coupling delay \cite{Sawicki2017}. This drift may be 
	used 
	in control schemes for stabilization of the chimera's position  
	\cite{Bick2015,Omelchenko2016,Omelchenko2018Tweezer}. The drastic 
	difference of 
	our model is that the core motion is random-like, but it does not vanish as 
	the 
	network size grows. Therefore we consider itinerant  chimeras reported 
	herein 
	as a novel dynamical regime observed for the first time.

	\section{Model}
	
	Our model is a network of phase oscillators with pulse coupling. The pulse 
	coupling scheme was used in order to speed up the numerical simulations by 
	using the effective reduction schemes \cite{Klinshov2017}. On the other 
	hand, 
	pulse-coupled oscillators are often seen as a conceptual model for 
	populations 
	of neurons. In the phase oscillator representation, neurons are 
	characterized 
	by their phase response  functions (PRCs) which may be calculated for any 
	neuronal model  \cite{Rinzel1998,Achuthan2009}. 
	
	A distinctive feature of our model is that the coupling weights are not  
	constant but rather evolve according to a certain plasticity rule. In the  
	context of neuronal networks, the coupling weights evolution corresponds 
	to  
	various plasticity mechanisms which change the strength of the synapses.  
	Recent studies have demonstrated the importance of the timing of individual 
	spikes in synaptic plasticity \cite{Gerstner1996,Morrison2008,Gilson2010}. 
	In 
	order to account for such spike-timing-dependent plasticity (STDP) in our 
	model 
	the dynamics of the coupling weights is phase-dependent. 
	
	Our network of $N$ identical all-to-all coupled oscillators is given by the 
	system 
	
	\begin{eqnarray}
	\frac{d\varphi_{j}}{dt} & = & \omega+\frac{1}{N}\sum_{k\neq 
		j}\kappa_{jk}\Gamma\left(\varphi_{j}\right)\sum_{t_{k}}\delta\left(t-t_{k}\right),\label{eq:1}\\
	\frac{d\kappa_{jk}}{dt} & = & 
	\varepsilon\left(-\kappa_{jk}+\Pi\left(\varphi_{j}\right)\sum_{t_{k}}\delta\left(t-t_{k}\right)\right).\label{eq:2}
	\end{eqnarray}
	
	Here, $\varphi_{j}\in[0;2\pi]$ is the $j$-th oscillator's phase, 
	$\kappa_{jk}$
	is the strength of the connection from $k$-th to $j$-th oscillator
	\cite{kappa_jk}, $\Gamma(\varphi)$ is the phase response curve,
	$\varepsilon$ is a (small) parameter controlling the adaptation rate,
	while function $\Pi(\varphi)$ defines the plasticity rule (see Fig. 
	\ref{fig:sketch}). In the
	absence of coupling, each oscillator has the same native frequency
	$\omega=1$, and its phase grows uniformly. When the phase reaches
	$2\pi$, it resets to zero, and the oscillator emits a pulse. 
	The coupling is pulse-like and described by the double sum in (\ref{eq:1}). 
	The first sum runs over all oscillators $k\neq j$, while the second sum 
	runs 
	over all moments $t_{k}$ when the $k$-th oscillator produces pulses. Each
	pulse is instantly received by the $j$-th oscillator and causes the
	latter's momentary phase shift 
	$\Delta\varphi_{j}=\kappa_{jk}\Gamma\left(\varphi_{j}\right)$.
	We take the phase response curve in the form 
	$\Gamma(\varphi)=-\sin(\varphi+\alpha)$, where $\alpha$ is the coupling 
	phase 
	lag. 
	
	In the absence of pulses, the coupling coefficients $\kappa_{jk}$
	relax to zero with the rate defined by $\varepsilon$. Each pulse
	produced by oscillator $k$ leads to momentary change of its connections
	to all other oscillators so that $\kappa_{jk}$ changes by 
	$\Delta\kappa_{jk}=\Pi(\varphi_j)$. The plasticity rule is given by the 
	function
	$\Pi(\varphi)=\sin(\varphi+\beta)$, where $\beta$ allows to control
	various modalities. For example, $\beta=\pi$ gives rise to an STDP-like
	plasticity rule, while $\beta=3\pi/2$ qualitatively represents the
	Hebbian learning rule \cite{Aoki2009,Aoki2011}.
	
	\begin{figure}[t]
		\includegraphics[width=85mm]{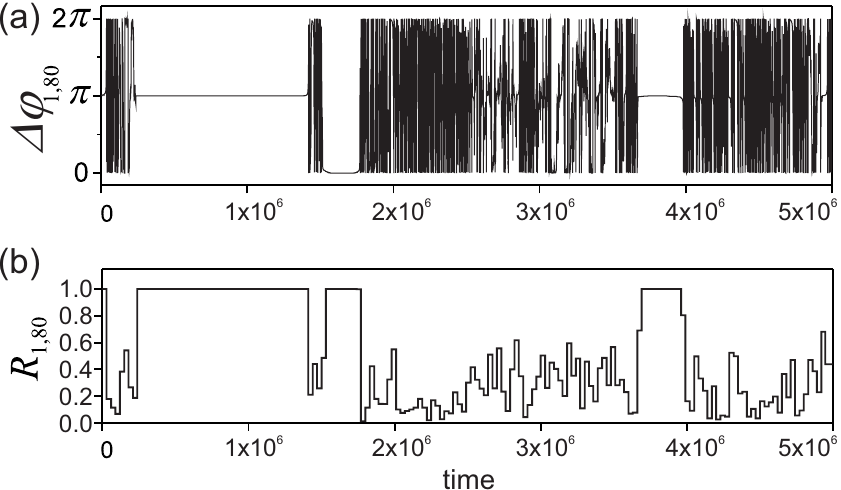}
		
		\caption{\label{fig:phasedyn}The dynamics of a randomly chosen pair of 
			oscillators. (a) The phase lag between the two oscillators 
			${\Delta\varphi_{1,80}\equiv\varphi_1-\varphi_{80}\mbox{ 
			mod}2\pi}$. (b) 
			The 
			transient 
			degree of synchrony between the same oscillators. The network size
			$N=200$, the parameters $\varepsilon=0.01$, $\alpha=1.4$ and 
			$\beta=4.94$.}
	\end{figure}
	
	\section{Results}
	
	For the rest of the paper, we use the parameter values $\varepsilon=0.01$,
	$\alpha=1.4$ and $\beta=4.94$ unless otherwise is stated. We observe the 
	dynamics of the network starting from random initial conditions. The 
	initial 
	phases are drawn from a uniform distribution $\varphi\in[0,2\pi]$, the 
	coupling 
	coefficient from a uniform distribution $\kappa\in[-1,1]$. 
	
	While observing the network collective dynamics, our attention was drawn by 
	a 
	peculiar regime which to the best of our knowledge has not been reported 
	before. We first noticed this regime when observed the temporal dynamics of 
	phase lags between different oscillators. For certain parameters, these 
	lags 
	demonstrated intermittent behavior: the two oscillators alternated between 
	the 
	periods of phase locking and incoherence. This behavior is illustrated in 
	Fig. 
	\ref{fig:phasedyn}a for a randomly selected pair of oscillators, and it is 
	very 
	similar for all other pairs. 
	
	In order to gain sight of a broader picture on the whole network scale
	we calculated the transient degree of synchrony between the oscillators
	defined as follows:
	
	\begin{equation}\label{eq:R}
	R_{jk}(t)=\frac{1}{\Delta}\left|\int_{t}^{t+\Delta}e^{i\left[\varphi_{j}(t)-\varphi_{k}(t)\right]}dt\right|.
	\end{equation}
	
	Here, $t$ is the current time, and $\Delta$ is a (sufficiently large) time 
	window. In order to capture the intermittent behavior described above, 
	$\Delta$ 
	must be much larger than the native period of the oscillators, but much 
	smaller 
	than the typical duration of the coherence/incoherence episodes. Further we 
	usd 
	$\Delta=3000$, but the results do not significantly change for other
	values of $\Delta$ in a wide range. Figure \ref{fig:phasedyn}b shows the 
	evolution of the transient degree of synchrony between the two oscillators 
	whose
	dynamics is depicted in Fig. \ref{fig:phasedyn}a. It is close to one while 
	the phases of the two oscillators are locked, and smaller than
	one while they drift apart. 
	
	\begin{figure*}[!t]
		\includegraphics[width=130mm]{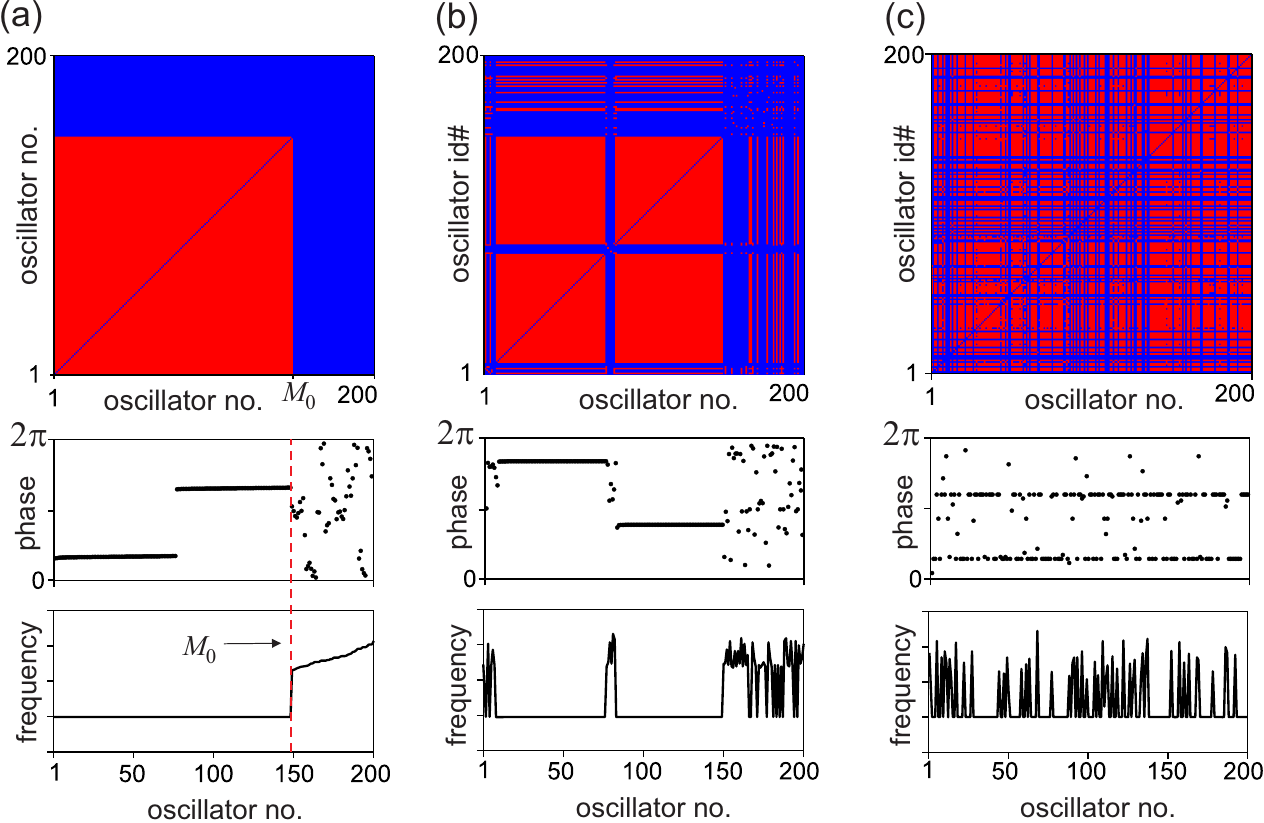}
		
		\caption{\label{fig:syncmatr}The network states at subsequent time 
		moments: 
			(a) 
			at $t=t_{0}=5\times10^5$, after the indexes renumbering; (b) at 
			$t=t_{1}=6.5\times10^5$; (c) at $t=t_{2}=8.5\times10^5$. In the
			upper panels, the matrix of the transient degrees of synchrony 
			$R_{jk}$ 
			is 
			plotted. Red (light gray) corresponds to strong synchrony 
			$R_{jk}>R^*$, where $R^*=0.999$. Blue (dark gray) corresponds to 
			weak 
			synchrony $R_{jk}<R^*$. 
			In the middle and bottom panels, the phases and the mean 
			frequencies of 
			the 
			oscillators are presented. In (a), the core size $M_0$ is indicated 
			by 
			the red dashed line. The network size $N=200$, the parameters 
			$\varepsilon=0.01$, $\alpha=1.4$ and 
			$\beta=4.94$.}
	\end{figure*}

	We analyzed the degree of synchrony of all the oscillator pairs across the 
	network depending on time. The resulting matrices $R_jk$ are presented in 
	Fig. 
	\ref{fig:syncmatr} along with the snapshots of the phases and mean 
	frequencies. 
	The mean frequency  of each oscillator was calculated along the time 
	interval 
	$\Delta$. The major finding is that the oscillators split into two domains. 
	The 
	first domain demonstrates strong synchronization which is manifested by the 
	degree of synchrony close to one. The oscillators of the second domain show 
	low 
	synchrony with those from the first domain and also among each other. In 
	order 
	to reveal this splitting we renumbered the oscillators according to their 
	degree of synchrony at the particular time moment $t=t_{0}=5\times 10^5$. 
	Then, 
	the coherent domain consists of the oscillators with indexes from $1$ to 
	$M_0$, 
	and the incoherent one of the oscillators with indexes from $M_0+1$ to $N$, 
	where $M_0=148$ is the size of the coherent domain at $t_{0}$. The state of 
	the 
	network with renumbered indexes is illustrated in Fig. \ref{fig:syncmatr}a. 
	One 
	sees that the coherent domain consists of two anti-phase clusters, while 
	the 
	incoherent domain has a broad distribution of phases. The frequencies of 
	the 
	oscillators from the coherent domain are equal, while those of the 
	incoherent 
	domain are widely distributed. This frequency profile is typical for 
	chimera 
	states.
	
	\begin{figure}[!t]
		
		\includegraphics[width=85mm]{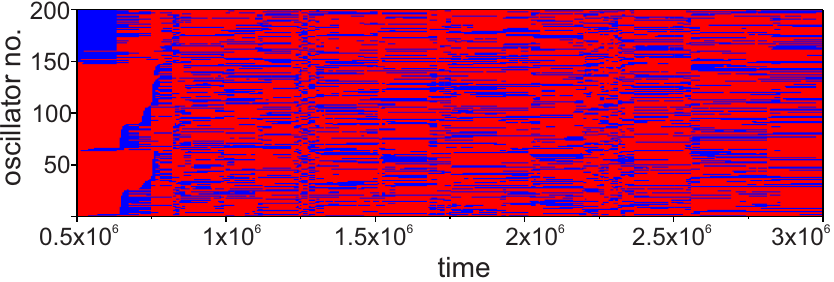}
		
		\caption{\label{fig:wandering}The evolution of the coherent and 
		incoherent domains of the coherent of the itinerant  chimera. Red 
		(light gray) corresponds to the coherent, blue (dark gray) to the 
		incoherent domain.			The 
			network 
			size $N=200$, the parameters 		$\varepsilon=0.01$, 
			$\alpha=1.4$ and 
			$\beta=4.94$.}

		\begin{center}	
			\includegraphics[width=50mm]{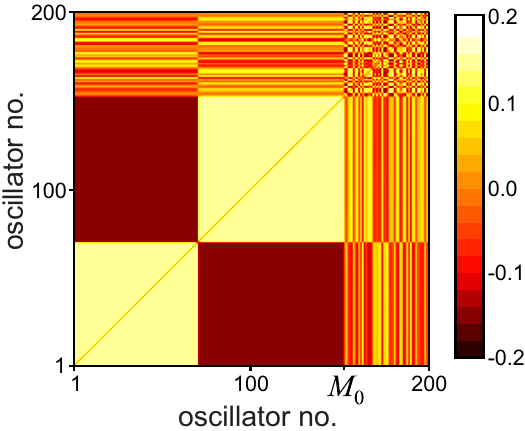}
		\end{center}
		
		\caption{\label{fig:coupling}Coupling matrix $\kappa_{jk}$ at the 
		moment 
			$t=t_{0}$ after the oscillators renumbering.  The network size 
			$N=200$, the 
			parameters $\varepsilon=0.01$, $\alpha=1.4$ and 
			$\beta=4.94$.}

		\begin{center}	
			\includegraphics[width=65mm]{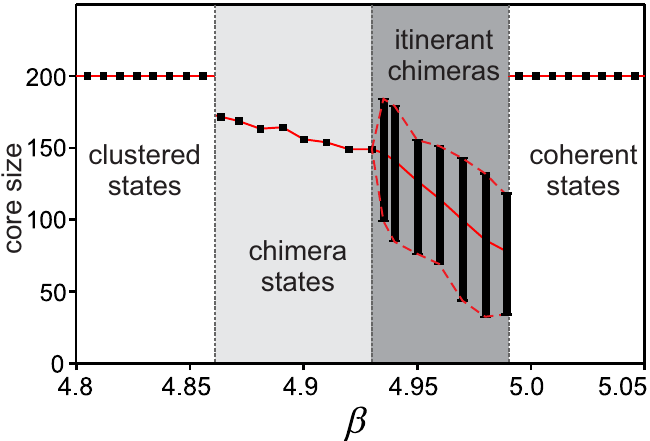}
		\end{center}	
		
		\caption{\label{fig:beta}Core size versus the parameter $\beta$. Black 
		bars 
			denote the points where the simulations were performed. For 
			itinerant  
			chimeras, the mean core size is plotted by solid line, minimal and 
			maximal 
			values by dashed lines. The network size $N=200$, the resting 
			parameters 
			$\varepsilon=0.01$,
			$\alpha=1.4$.}

	\end{figure}
	
	The splitting of the oscillators into the two domains, the coherent and the 
	incoherent ones, strongly resembles a chimera state. However, there is a 
	drastic difference between the classical chimeras and the regime that we 
	observe. In order to trace it we fixed the oscillator indexes and observed 
	the 
	long-term evolution of the network. Then we noticed that the composition of 
	the 
	coherent and the incoherent domains was volatile, meaning that each 
	particular 
	oscillator spontaneously switched from one domain to another. In order to 
	demonstrate this volatility we illustrate the network states in subsequent 
	moments of time. In Fig. \ref{fig:syncmatr}b, the coherent and the 
	incoherent 
	domains are still present at $t=t_{1}=6.5\times 10^5$, but their 
	composition is 
	different compared to $t=t_0$. The oscillators which constitute the domains 
	are 
	not longer ordered but rather mixed across the network. This mixing goes 
	even 
	further in Fig. \ref{fig:syncmatr}c for $t=t_{2}=8.5\times 10^5$. 
	
	To better picture the process of mixing of the coherent and the incoherent  
	domains we studied the temporal evolution of their composition.  At each 
	time 
	moment we calculated the transient degrees of synchrony and determined the 
	domain attribute $u_j$ of each oscillator. The oscillator was attributed 
	belonging to the coherent domain with $u_j=1$ if it was strongly 
	synchronized 
	with some others, and to the incoherent one with $u_j=0$ if it has no 
	strong 
	synchrony with any others. The composition of the domains is plotted versus 
	time in Fig. \ref{fig:wandering}, red (light gray) corresponding to the 
	coherent, blue (dark gray) to the incoherent domain. A the time $t_0$ just 
	after the oscillators 
	renumbering 
	the domains are ordered. As the time passes the oscillators spontaneously  
	switch their domains. From the network viewpoint, this process corresponds 
	to 
	the volatility of the domains composition. The coherent domain, which is 
	often 
	called the chimera's core, does not stay in the same position but rather 
	moves 
	spontaneously across the network. This feature led us to adopt the name 
	``itinerant  chimera'' to the observed regime.

	The splitting of the oscillators into two domains is supported by the 
	sufficient
	structure of the coupling matrix which is depicted in Fig. 
	\ref{fig:coupling} 
	for $t=t_0$. Note that this moment corresponds to the network state 
	illustrated 
	in Fig. 2a when the oscillator indexes are renumbered so that the coherent 
	domain consists of oscillators $1,...,M_0$, and the incoherent domain of 
	oscillators $M_0+1,...,N$. Recall that the coherent domain consists from 
	two 
	anti-phase clusters. The oscillators within each synchronous cluster have 
	strong positive connections, while the two clusters have strong negative 
	connections between each other. These strong and structured connections are 
	the 
	reason for the synchrony within the coherent domain. At the same time, the 
	connections within the incoherent domain and between the two domains do not 
	show any structure, they may be either negative or positive as well as 
	strong 
	or weak. This diversity determines the lack of synchrony within the 
	incoherent
	domain. Note that the structure of the coupling matrix is not prescribed
	but rather emerges from the random initial conditions due to the network
	adaptivity. During the further network evolution when the core composition 
	changes, the coupling matrix changes as well, but its basic features 
	preserve: 
	the connections within the core are strong and well structured, while the 
	resting connections do not show any structure.
	
	The itinerant chimeras are robust patterns which persist under the 
	variation of 
	the system parameters. In order to prove that we changed the parameters 
	$\alpha$, $\beta$ and $\varepsilon$ and analyzed the observed behavior 
	patterns. The results are presented in Fig.  \ref{fig:beta} where the core 
	size 
	$M$ is plotted versus the parameter $\beta$ (the data for other parameters 
	variation is not shown). The core size is calculated as the number of the 
	synchronized oscillators $M=\sum_j u_j$. For $\beta<4.86$ the oscillators 
	split 
	into several clusters of synchrony, therefore the ``core'' occupies the 
	whole 
	network and $M=N$. For $4.86<\beta<4.93$ classical chimeras are observed 
	for 
	which the composition and the size of the core do not change with time. The 
	itinerant  chimeras are observed for $4.93<\beta<4.99$, and in this 
	parameter 
	interval not only the constitution, but also the size of the core changes 
	with 
	time. For $\beta>4.99$ the system undergoes a transition to the coherent 
	state.
	
\begin{figure}[t]

		\includegraphics[width=85mm]{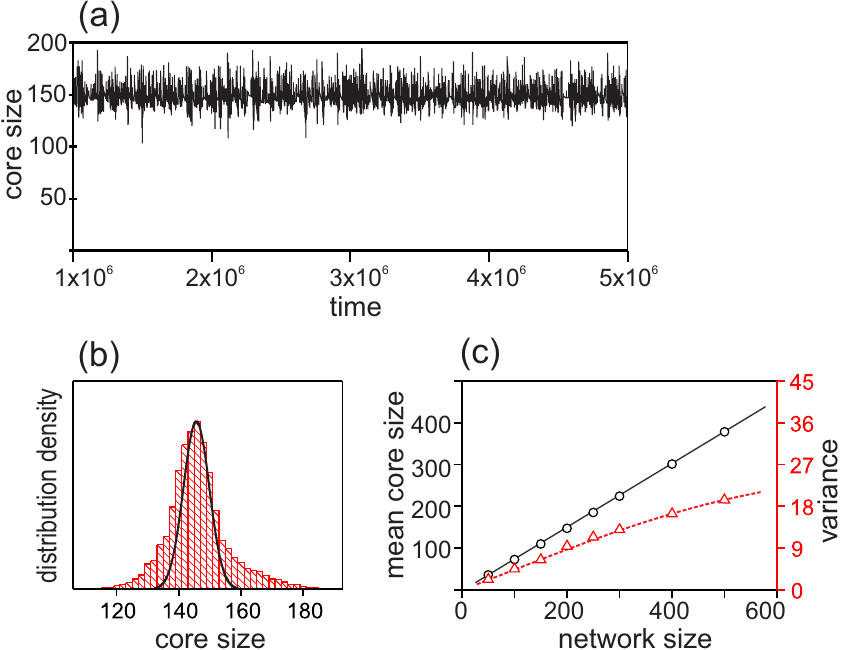}
		
		\caption{\label{fig:size} (a) The chimera core size versus time 
			for $N=200$. (b) Distribution of the core size for $N=200$.
			Black solid line corresponds to the binomial distribution. (c) The
			mean (black solid line) and the variance (red dashed line) of the 
			core size versus the network size. The
			parameters $\varepsilon=0.01$, $\alpha=1.4$ and 
			$\beta=4.94$.}
	\end{figure}

	Further we investigate in more details the traveling of the core and 
	demonstrate that it is not only a finite-size effect, but a keynote feature 
	of 
	the network dynamics which preserves even for large number of nodes. The 
	dynamics of the core size is illustrated in Fig. \ref{fig:size}a, it 
	demonstrates pronounced fluctuations around the mean. The size fluctuations 
	suggest random-like transitions of the oscillators between the domains. 
	These 
	transitions appear to be positively correlated meaning that the 
	oscillators  
	tend to switch their domain in groups. In order to illustrate that we plot 
	the 
	distribution of the chimera core size $M$ observed in a long time interval 
	in 
	Fig. \ref{fig:size}b. For the independent random-like switching of 
	individual 
	oscillators the core size would have binomial distribution 
	$M\sim\text{B}(N,p)$, where $p$ is the fraction of the oscillators 
	belonging to 
	the core. However, the  obtained distribution is much wider and has much 
	heavier tails suggesting concurrent transitions of large groups of 
	oscillators. 
	
	In order to study the scaling of the itinerant chimeras  we plot the mean 
	and 
	the variance of the core size $M$ versus the network size $N$ in Fig. 
	\ref{fig:size}c. The mean core size $\left\langle M\right\rangle$ grows 
	linearly with the network size suggesting the constant ratio between the 
	coherent and the incoherent domains. The fraction of the oscillators in the 
	core may be estimated as $p=\left\langle M\right\rangle /N\approx0.72$. The 
	variance $\sigma_{M}=\sqrt{\left\langle M^{2}\right\rangle -\left\langle 
		M\right\rangle ^{2}}$ grows sub-linearly, however, it is much larger 
		than 
	predicted by the binomial distribution. The wide distribution of the core 
	size 
	corroborates that the core volatility manifests itself on the macroscopic 
	level, not only as a finite-size effect.
	
	To prove random-like character of the oscillators switching between the 
	domains, we introduce the autocorrelation function of the core composition 
	defined as
	
	\begin{equation}\label{eq:A}
	A(\tau)=\frac{1}{\left\langle M\right\rangle 
	}\lim_{t\to\infty}\frac{1}{T}\int_{0}^{T}\sum_{j=1}^{N}u_{j}(t)u_{j}(t+\tau)dt.
	\end{equation}
	
\begin{figure}[t!]
		\includegraphics[width=85mm]{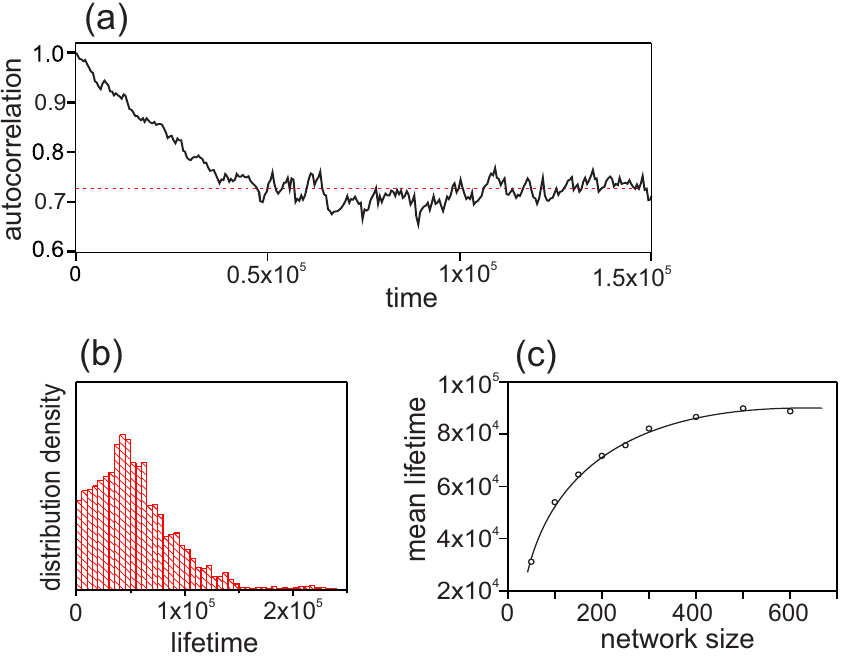}
		
		\caption{\label{fig:lifetime}(a) Autocorrelation function of the 
			core composition versus time for $N=200$. The horizontal dashed 
			line 
			corresponds to  $A=p$. (b) Distribution of the nodes
			lifetimes in the core for $N=200$. (c) The mean lifetime versus the
			network size. The parameters $\varepsilon=0.01$, $\alpha=1.4$ and 
			$\beta=4.94$.}
	\end{figure}

	Here, the sum under the integral is nothing else but the number of the 
	oscillators which belong to the core at the both time moments $t$ and 
	$t+\tau$. 
	The function $A(\tau)$ gives the mean fraction of the oscillators which 
	stay in 
	the core (or return back to it) in time $\tau$. The autocorrelation 
	function 
	$A(\tau)$ is plotted in Fig. \ref{fig:lifetime}a. It equals unity at  
	$\tau=0$ 
	and falls to $A\approx p$ at $\tau\sim5\times10^4$  which value corresponds 
	to 
	the fraction of units shared  by two randomly selected sets. This means 
	that 
	the network memory about the core composition fades completely in this 
	time, 
	the core effectively spreads across the network, and its composition 
	becomes 
	unpredictable. The correlation decay is a clear sign of the chaotic 
	dynamics, 
	which we confirmed by the calculation of the largest Lyapunov exponent 
	$\lambda=0.0045$. Note however, that the inverse time $\lambda^{-1}=222$ is 
	much shorter than the typical time of the core spreading. 
	
	Another way to estimate the rate of the core traveling is to compute the 
	lifetimes of individual oscillators in the core.  The distribution of the 
	lifetimes is shown in Fig. \ref{fig:lifetime}b, it is a broad unimodal 
	distribution with the average of about $5\times10^4$ which roughly 
	corresponds 
	to the result from Fig. \ref{fig:lifetime}a. The scaling behavior of the 
	mean 
	lifetime is illustrated in Fig. \ref{fig:lifetime}c. Although the lifetime 
	grows with  the network size, this growth is relatively slow and tends to 
	saturate, in sharp contrast with the lifetime of classic chimeras which was 
	shown to increase exponentially \cite{Wolfrum2011PRE}. Thus, the finite 
	speed 
	traveling preserves even for large networks.
	
	\section{Conclusions}
	
	We have studied a new type of chimera-like behavior observed in networks of 
	oscillators with adaptive coupling. Similarly with classical chimeras, the 
	oscillators split into two domains, the coherent and the incoherent ones. 
	However, the drastic difference is that the composition of the coherent and 
	incoherent domains changes with time. The oscillators spontaneously switch 
	their domain which results in traveling of the chimera's core across the 
	network. This process is  characterized by fading memory, meaning that the 
	network forgets the composition of the core in a finite time. The lifetime 
	of 
	the core grows slowly with the network size suggesting that the core 
	volatility 
	is not a finite-size effect but rather an intrinsic feature of the network 
	collective dynamics.
	
	Our system may also demonstrate other collective behaviors depending on the 
	parameters $\alpha$ and $\beta$. In particular, we observed classical 
	chimera 
	states and the emergence of clustered states similar to those described 
	earlier 
	for continuous coupling \cite{Kasatkin2017}. However, we conjecture that 
	the 
	pulse nature of coupling together with its adaptivity was crucial for the 
	emergence of itinerant  chimeras. Provided that the major motivation for 
	the 
	model comes from neuroscience, it would be intriguing to search for similar 
	dynamical regimes in biologically plausible setups and explore their 
	possible 
	role in collective dynamics of neuronal populations.
	
	The authors are grateful to Dr. Anna Zakharova and Dr. Christian Bick
	for many helpful discussions. The work was supported by the Russian
	Foundation for Basic Research (Projects No. 17-02-00904, 17-02-00874, 
	18-02-00406 and 18-29-10040).


\begin{thebibliography}{10}
\bibitem{kappa_jk}Note that the matrix $\kappa_{jk}$ encodes both
		the weights and topology of the connections: even if we start from
		all-to-all coupling, in principle some connections may weaken with
		time and extinct due to the plasticity. Note also that we do not allow 
		for 
		self-coupling.
\end{thebibliography}
\end{document}